# Lattice-pseudospin and spin-valley polarizations in dual ferromagnetic-gated silicene junction

**Peerasak Chantngarm [a], Kou Yamada [a], Bumned Soodchomshom [b,*]**

[a] Domain of Mechanical Science and Technology,

Graduate School of Science and Technology, Gunma University, Gunma, Japan

[b] Department of Physics, Faculty of Science, Kasetsart University Bangkok 10900, Thailand

Corresponding author:

Email: Bumned@hotmail.com, fscibns@ku.ac.th (B. Soodchomshom)

**Abstract:**

We study spin-valley and lattice-pseudo spin currents in a dual ferromagnetic-gated silicene-based junction. Silicene has buckled atomic structure which allows us to take sublattice-dependent ferromagnetism into account in the investigation. One of the study results show that transmission at the junctions exhibits anisotropic property only in anti-parallel cases. Interestingly, the studied junctions can be switched from a pure spin-polarizer to a pure valley-polarizer by reversing directions of exchange fields in the parallel junctions. The perfect control of spin-valley currents can be done only in the parallel cases and its resolution can be enhanced by increasing gate potential between the ferromagnetic barriers. The asymmetric barriers of anti-parallel junction is found to destroy both spin and valley filtering effects and yield a novel result, pure sub-lattice pseudo-spin polarization. The current in the anti-parallel junctions can be controlled to flow solely in either A or B sub-lattice, saying that the controllable lattice current in silicene is created in double ferromagnetic-gated junction. Our work reveals the potential of dual ferromagnetic-gated silicene junction which may be possible for applications in spin-valleytronics and lattice-pseudospintronics.

# 1. INTRODUCTION

After graphene [1], many other two-dimensional (2D) materials [2] are theoretically discovered, such as germanene (Ge) [3], phosphorene (P) [4], tinene (Sn) [5] and silicene (Si) [6]. Silicene has atomic structure akin to graphene but with buckling, so some electronic properties are different from those in graphene, such as large spin-orbit interaction. At first, silicene seemed to be elusive materials due to its very sensitive surface. However, after experimental success [7-8], it is considered to be one of the most promising 2D materials in electronics applications due to the accumulation of silicon-related technology and knowhow in semiconductor industry. The recent study on field-effect transistors operating at room temperature made from silicene [9] is particularly considered to be a big leap in this area. To enhance the capability of conventional electronics devices based on charge degree of freedom, spintronic devices based on spin degree of freedom have become more important [10]. More recently, valley degree of freedom based on two inquivalent Dirac points at $k$ and $k'$ has also attracted interests [11] as a pathway towards quantum computing. The presence of large spin-orbit interaction and buckled atomic structure lead silicene to be a candidate for these growing fields of spintronics [12] and valleytronics [13].

Silicene and other monolayers of honeycomb-lattice atoms are parts of Dirac materials [14], which also include topological insulators and high-temperature d-wave superconductors. The properties that make these group of materials unique and exciting are from the fact the low-energy electrons in these condensed matter systems obey Dirac equation in stead of Schrödinger equation. The massless Dirac fermions in the systems give rise to many interesting phenomena, for example, integer quantum Hall effect (QHE) [15], Klein paradox [16], fractional quantum Hall effect (FQHE) [17]. With better understanding in the properties of this group of materials including silicene, we expect to make tremendous impact on the higher computing power and other technological areas.

Although silicene and its more analyzed predecessor, graphene, are the same 2D Dirac materials, there are few significant differences between them. For example, silicene has buckled honeycomb lattice structure which in turn allows Dirac electron mass to be manipulated by electric field [18-19]. Silicene also has stronger spin-orbit coupling which gives rise to the spin-valley coupling [20]. There have been various theoretical studies in spin-valley transport at silicene junction, which helps the advancement in this area. The topics of those investigations are, for example, the electric field condition for the fully valley and spin polarized transports [21], the mechanism of magnetism opening different spin dependent band gaps at k and k' points which results in spin and valley polarized transports [20], ballistic transport through silicene FM junctions [22], and the transmission probability and valley conductance relating to the local electric field and exchange field [23]. Other studies have also been made in electron transport of silicene based spintronics and valleytronics devices [24-30], where more attention is attracted recently. The topics of studies are such as spin filter and spin-valley filter [24-26], spin thermoelectric properties [27], spin-polarized transport in a dual-gated silicene system without exchange field [28], epitaxial growth of multilayer silicene [29] and using electric and exchange fields to tune the plasmonic response of the electron gas in silicene [30]. However, more analysis must be done to completely understand the properties of this material, not only spin and valley currents, for being used in real-world applications. The latice-pseudospin currents are also applicable for the so-called "lattice-pseudospintronics", devices that control currents to flow in either A or B sublattice atomic structure.

In this paper, we study the spin-valley current and lattice-pseudospin current in silicene-based normal(NM)/ ferromagnetic(FM)/ normal(NM)/ ferromagnetic(FM)/ normal (NM) junction, effected by the presence of ferromagnetic dual gated barriers. The wave equation of the carriers is described by the low energy tight-binding-based Hamiltonian [13, 24]. In this work, we show that the studied junction may be anisotropic transport property and destroy spin-valley filtering, in the anti-parallel type. By focusing on the anisotropic properties of the device and the characteristics of spin and valley currents, we propose a new way to control spin-valley currents and lattice-pseudospin currents in silicene with dual magnetic gates for both parallel and anti-parallel junctions. The potential of double magnetic-gated silicene junction would be revealed for applications in spin-valley-current and lattice-pseudo-spin current based devices.

## 2. MODEL

The schematic model of double-barrier silicene-based structure, NM1 / FM1 / NM2 / FM2 / NM3, is shown in Fig.1. Each of the magnetic barriers, FM1 and FM2, has length d with distance L separating from each other. The magnetic barriers are induced into ferromagnetism by a pair of magnetic insulators on both sides of the silicene sheet. The exchange energies, which are induced by the magnetic insulators into A- and B-sublattices, are designated as $h_{1A}$ and $h_{1B}$ at FM1, while they are designated as $h_{2A}$ and $h_{2B}$ at FM2, respectively.

Four junction types, two parallel junctions and two anti-parallel junctions, are used in this investigation as shown in Fig. 2. In the parallel junctions (P-1 and P-2), the exchange fields at FM1 and FM2 have the same direction on each side of silicene layer, while the direction is opposite in anti-parallel junctions (AP-1 and AP-2). The chemical potential μ is induced by potential μ/e from the top and the bottom gates. Due to the buckling structure, silicene sheet has a perpendicular distance between A- and B-sublattices of 2D = 0.43 Å [13]. The controllable electric field $E_z$ is applied into the barrier-regions, FM1 and FM2, and is perpendicular to silicene sheet. The gate potential U/e is also applied at the silicene NM2 layer.

Based on the behavior of non-interacting electrons in NM1, NM2 and NM3 layers described by Kane-Mele model [31], as well as the behavior of electrons in silicene under the influence of the gate potential, electric field, and exchange energy in FM1 and FM2 [13], the tight-binding Hamiltonians and low-energy effective Hamiltonians used to describe the motion of electrons in A- and B-sublattices were analyzed [24]. The effect of Rashba interaction is very small comparing with the other terms at low energy [13, 21]. Therefore, when $k$ ($k'$) valley is represented by $\eta = 1$ (-1), and spin ↑ (↓) is represented by σ = 1 (-1) respectively, the wave equation with excited energy E is obtained by [13, 21, 24]

$$\hat{H}_{\eta\sigma}\psi_{\eta\sigma} = E\psi_{\eta\sigma} \,. \qquad (1)$$

The Hamiltonian here acts on the spin-valley-dependent "lattice-pseudospinnor field" $\psi_{\eta\sigma} = \begin{pmatrix} \Psi_{A,\eta\sigma} \\ \Psi_{B,\eta\sigma} \end{pmatrix}$, where $\Psi_{A,\eta\sigma}$ and $\Psi_{B,\eta\sigma}$ are wave functions of electrons with spin ↑, ↓ in k, k'-valleys at A- and B-sublattices, respectively. In NM1, NM2, and NM3, the Hamiltonian is defined as

$$\hat{H}_{\eta\sigma} = v_F(\hat{p}_x\tau^x - \eta\hat{p}_y\tau^y) - \Delta_{\eta\sigma}\tau^z, \qquad (2)$$

where $\hat{p}_x = -i\hbar\dfrac{\partial}{\partial x}$, $\hat{p}_y = -i\hbar\dfrac{\partial}{\partial y}$ and $\tau^x$, $\tau^y$, $\tau^z$ are elements of Pauli spin-operators used to represent "lattice pseudospin". The Fermi velocity near the Dirac point in this case is $v_F \cong 5.5 \times 10^5$ m/s [32]. $\Delta_{\eta\sigma} = \eta\sigma\Delta_{SO}$ represents the spin-valley-dependent energy gap in these regions, where $\Delta_{so}$ is spin-orbit interaction. In FM1, the Hamiltonian is defined as

$$\hat{H}_{\eta\sigma} = v_F(\hat{p}_x\tau^x - \eta\hat{p}_y\tau^y) - \Delta_{\eta\sigma 1}\tau^z - \mu_{\sigma 1}, \qquad (3)$$

where the spin-valley-dependent energy gap here is $\Delta_{\eta\sigma 1} = \eta\sigma\Delta_{SO} - \Delta_E + \sigma\Delta_{M1}$, with the electric field $\Delta_E = eDE_z$ and the exchange field-induced gap $\Delta_{M1} = (h_{1A} - h_{1B})/2$. The spin-dependent chemical potential is $\mu_{\sigma 1} = \mu + \sigma u_{M1}$, where $u_{M1} = (h_{1A} + h_{1B})/2$, since the chemical potential in the barrier is spin-dependent relating to the exchange field. In FM2, the Hamiltonian is similarly defined as

$$\hat{H}_{\eta\sigma} = v_F(\hat{p}_x \tau^x - \eta \hat{p}_y \tau^y) - \Delta_{\eta\sigma2}\tau^z - \mu_{\sigma2}, \qquad (4)$$

where $\Delta_{\eta\sigma2} = \eta\sigma\Delta_{SO} - \Delta_E + \sigma\Delta_{M2}$, $\mu_{\sigma2} = \mu + \sigma u_{M2}$, and $u_{M2} = (h_{2A} + h_{2B})/2$, with the exchange field-induced gap $\Delta_{M2} = (h_{2A} - h_{2B})/2$.

## 3. SCATTERING PROCESS

The Hamiltonian in Eq. (1) is used to describe the motion of electrons in this system, where the spin-valley currents are flowing along the x-direction. The wave functions in the NM1, FM1, NM2, FM2, and NM3 are respectively given as

$$\psi_{NM1} = \left[ \begin{pmatrix} 1 \\ A_{\eta\sigma}e^{-i\eta\theta} \end{pmatrix} e^{ik_x x} + r_{\eta\sigma} \begin{pmatrix} 1 \\ -A_{\eta\sigma}e^{i\eta\theta} \end{pmatrix} e^{-ik_x x} \right] e^{ik_{//}y},$$

$$\psi_{FM1} = \left[ a_{\eta\sigma} \begin{pmatrix} 1 \\ B_{\eta\sigma}e^{-i\eta\alpha_1} \end{pmatrix} e^{il_x x} + b_{\eta\sigma} \begin{pmatrix} 1 \\ -B_{\eta\sigma}e^{i\eta\alpha_1} \end{pmatrix} e^{-il_x x} \right] e^{ik_{//}y},$$

$$\psi_{NM2} = \left[ g_{\eta\sigma} \begin{pmatrix} 1 \\ M_{\eta\sigma}e^{-i\eta\beta} \end{pmatrix} e^{im_x x} + f_{\eta\sigma} \begin{pmatrix} 1 \\ -M_{\eta\sigma}e^{i\eta\beta} \end{pmatrix} e^{-im_x x} \right] e^{ik_{//}y},$$

$$\psi_{FM2} = \left[ p_{\eta\sigma} \begin{pmatrix} 1 \\ N_{\eta\sigma}e^{-i\eta\alpha_2} \end{pmatrix} e^{in_x x} + q_{\eta\sigma} \begin{pmatrix} 1 \\ -N_{\eta\sigma}e^{i\eta\alpha_2} \end{pmatrix} e^{-in_x x} \right] e^{ik_{//}y},$$

$$\psi_{NM3} = \left[ t_{\eta\sigma} \begin{pmatrix} 1 \\ A_{\eta\sigma}e^{-i\eta\theta} \end{pmatrix} e^{ik_x x} \right] e^{ik_{//}y},$$

where

$$A_{\eta\sigma} = \frac{E + \eta\sigma\Delta_{SO}}{\sqrt{E^2 - \Delta_{SO}^2}}, \ B_{\eta\sigma} = \frac{E + \mu_{\sigma1} + \Delta_{\eta\sigma1}}{\sqrt{(E + \mu_{\sigma1})^2 - \Delta_{\eta\sigma1}^2}},$$

$$M_{\eta\sigma} = \frac{E + U + \eta\sigma\Delta_{SO}}{\sqrt{(E + U)^2 - \Delta_{SO}^2}}, \ N_{\eta\sigma} = \frac{E + \mu_{\sigma2} + \Delta_{\eta\sigma2}}{\sqrt{(E + \mu_{\sigma2})^2 - \Delta_{\eta\sigma2}^2}}. \qquad (5)$$

The wave vectors in the x-direction of electron in NM and FM region are given by

$$k_x = \frac{\sqrt{E^2 - \Delta_{SO}^2}\cos\theta}{\hbar v_F}, \ l_x = \frac{\sqrt{(E + \mu_{\sigma1})^2 - \Delta_{\eta\sigma1}^2}\cos\alpha_1}{\hbar v_F}, \ m_x = \frac{\sqrt{(E + U)^2 - \Delta_{SO}^2}\cos\beta}{\hbar v_F},$$

$$n_x = \frac{\sqrt{(E + \mu_{\sigma2})^2 - \Delta_{\eta\sigma2}^2}\cos\alpha_2}{\hbar v_F} \qquad (6)$$

The incident angle $\alpha_1$ at the FM1/NM2 barrier, the incident angle $\beta$ at the NM2/FM2 barrier and the incident angle $\alpha_2$ at the FM2/NM3 barrier can be calculated via the conservation component in the y-direction as given by

$$k_{//} = \frac{\sqrt{E^2 - \Delta_{SO}^2} \sin\theta}{\hbar v_F} = \frac{\sqrt{(E + \mu_{\sigma 1})^2 - \Delta_{\eta\sigma 1}^2} \sin\alpha_1}{\hbar v_F} = \frac{\sqrt{(E + U)^2 - \Delta_{SO}^2} \sin\beta}{\hbar v_F}$$

$$= \frac{\sqrt{(E + \mu_{\sigma 2})^2 - \Delta_{\eta\sigma 2}^2} \sin\alpha_2}{\hbar v_F}, \qquad (7)$$

where $\theta$ is the incident angle at the NM1/FM1 barrier. The coefficients $r_{\eta\sigma}$, $a_{\eta\sigma}$, $b_{\eta\sigma}$, $g_{\eta\sigma}$, $f_{\eta\sigma}$, $p_{\eta\sigma}$, $q_{\eta\sigma}$, $t_{\eta\sigma}$ can be calculated through the boundary conditions at the interfaces where

$$\Psi_{NM1}(0) = \Psi_{FM1}(0), \; \Psi_{FM1}(d) = \Psi_{NM2}(d),$$

$$\Psi_{NM2}(d + L) = \Psi_{FM2}(d + L), \; \Psi_{FM2}(2d + L) = \Psi_{NM3}(2d + L), \qquad (8)$$

with $r_{\eta\sigma}$ and $t_{\eta\sigma}$ represent reflection and transmission coefficients, respectively. The transmission probability amplitude $T_{\eta\sigma}$ is calculated via the formula $T_{\eta\sigma} = J_t / J_{in} = |t_{\eta\sigma}|^2$, where $J_t$ and $J_{in}$ are current densities of transmitted electrons and injected electrons, respectively.

## 4. TRANSPORT FORMULAE

Using the standard Landauer's formalism [33], the spin-valley conductance at zero temperature in the ballistic regime can be calculated by integrating overall the incident angles [15] as shown below

$$G_{\eta\sigma} = \frac{e^2}{h} N(E) |t_{\eta\sigma}|^2 = G_0 \frac{\sqrt{E^2 - \Delta_{SO}^2}}{|E|} \int_{-\pi/2}^{\pi/2} \frac{1}{8} d\theta \cos(\theta) T_{\eta\sigma}. \qquad (9)$$

Here, $G_0 = \frac{4e^2}{h} N_0(E)$ is unit conductance, where $N_0(E) = \frac{W}{\pi\hbar v_F} |E|$ is density of state at transport channel in silicene excluding spin-orbit interaction effect. W represents width of silicene sheet, h represents Planck's constant, and $N(E) = \frac{W}{\pi\hbar v_F} \sqrt{E^2 - \Delta_{SO}^2}$ represents density of state at the transport channel of normal silicene junction. Furthermore, the total conductance $G_T$ can be calculated using the summation of all spin-valley conductance $G_T = G_{k\uparrow} + G_{k\downarrow} + G_{k'\uparrow} + G_{k'\downarrow}$, where k is represented by $\eta = 1$, k' by $\eta = -1$, spin $\uparrow$ by $\sigma = 1$, and spin ↓ by σ = -1. The spin polarization (SP) and valley polarization (VP) of the junction are defined as

$$SP(\%) = \frac{\left(G_{k\uparrow} + G_{k'\uparrow}\right) - \left(G_{k\downarrow} + G_{k'\downarrow}\right)}{G_T} \times 100 \,,$$

and

$$VP(\%) = \frac{\left(G_{k\uparrow} + G_{k\downarrow}\right) - \left(G_{k'\uparrow} + G_{k'\downarrow}\right)}{G_T} \times 100 \,. \tag{10}$$

In this section, we will introduce lattice-pseudospin polarization, by considering the expectation value of lattice-pseudospin [35, 36] in NM1 or NM3 region using the formula in quantum mechanics $\langle \vec{s} \rangle_{\eta\sigma} = \left\langle \psi_{\eta\sigma} \left| \hat{\vec{s}} \right| \psi_{\eta\sigma} \right\rangle$, where $\psi_{\eta\sigma} = \begin{pmatrix} \psi_{A,\eta\sigma} \\ \psi_{B,\eta\sigma} \end{pmatrix}$ is the normalized wave function of electron in NM regions. A and B represent ⇑ and ⇓ transverse lattice-pseudospin states respectively, which are Eigenstate of $\hat{s}_z = \frac{\hbar}{2}\tau^z$. The lattice-pseudospin operator is defined as $\hat{\vec{s}} = \frac{\hbar}{2}(\tau^x \hat{a}_x + \tau^y \hat{a}_y + \tau^z \hat{a}_z)$. Then we get

$$\langle \vec{s} \rangle_{\eta\sigma} = \frac{\hbar}{2}\left( \sqrt{1 - \left(\frac{\Delta_{so}}{E}\right)^2}\, \hat{a}_{xy} - \eta\sigma \frac{\Delta_{so}}{E} \hat{a}_z \right), \tag{11}$$

where $\hat{a}_{xy}$ is in-plane unit vector (xy-plane) and $\hat{a}_z$ is out-of-plane unit vector (z-direction). We will see that when $E \to \Delta_{SO}$, then

$$\langle \vec{s} \rangle_{\eta\sigma} \to -\eta\sigma \frac{\hbar}{2}\hat{a}_z \,. \tag{12}$$

Here, we can classify electrons in silicene into two lattice-pseudospin groups with regard to the limit $E \to \Delta_{so}$, as given by

$$\langle \vec{s} \rangle_{k\uparrow} = \langle \vec{s} \rangle_{k'\downarrow} = -\frac{\hbar}{2}\hat{a}_z \equiv \left| B - sublattice \right\rangle \text{ or lattice-pseudospin down } \Downarrow \,,$$

and

$$\langle \vec{s} \rangle_{k\downarrow} = \langle \vec{s} \rangle_{k'\uparrow} = +\frac{\hbar}{2}\hat{a}_z \equiv \left| A - sublattice \right\rangle \text{ or lattice-pseudospin up } \Uparrow \,.$$

Using this result, we may get $\left| k \uparrow \right\rangle$ and $\left| k' \downarrow \right\rangle \equiv \left| \Downarrow \right\rangle$ while $\left| k \downarrow \right\rangle$ and $\left| k' \uparrow \right\rangle \equiv \left| \Uparrow \right\rangle$. Hence, the lattice-pseudospin polarization occurs when $E \to \Delta_{SO}$, which may give its definition as

$$LSP(\%) = \frac{G_\Uparrow - G_\Downarrow}{G_T} \times 100 = \frac{\left(G_{k\downarrow} + G_{k'\uparrow}\right) - \left(G_{k\uparrow} + G_{k'\downarrow}\right)}{G_T} \times 100, \quad (13)$$

where $G_\Uparrow = G_{k\downarrow} + G_{k'\uparrow}$ and $G_\Downarrow = G_{k\uparrow} + G_{k'\downarrow}$ are conductance of electrons with lattice-pseudospin up and down, respectively.

## 5. RESULT AND DISCUSSION

In our numerical calculation, the value of effective spin-orbit interaction is assumed to be $\Delta_{SO}$ =3.9 meV [28]. Fig. 2 shows the junction types used in this study, where parallel (P) junctions are defined as the conditions in Fig. 2 (a) when $h_{1A}$ = -$h_{1B}$ = $h_{2A}$ = -$h_{2B}$ = 5 meV (P-1) or in Fig. 2(c) when $h_{1A}$ = $h_{1B}$ = $h_{2A}$ = $h_{2B}$ = 5 meV (P-2), while the anti-parallel (AP) junctions are defined as the conditions in Fig. 2(b) when $h_{1A}$ = -$h_{1B}$ = -$h_{2A}$ = $h_{2B}$ = 5 meV (AP-1) or in Fig. 2(d) when $h_{1A}$ = $h_{1B}$ = -$h_{2A}$ = -$h_{2B}$ = 5 meV (AP-2), respectively. Here, the proximity-induced exchange energy of 5meV is used as proposed to induce ferromagnetism in graphene [34]. We focus on the transmission rates and conductance at the excited energy (E) approaching the spin-orbit energy gap $E \rightarrow \Delta_{SO}$ to show strong Dirac mass effect in NM regions [24]. The transverse lattice-pseudospin in NM regions approaches $s_z \rightarrow \pm \hbar/2$ in this condition [35, 36], saying that in-plane pseudospin is very small (also see Eq.12). Throughout this numerical study, we set L=25 nm, d=25 nm, h=5 meV, and E=4 meV.

Before showing the result, we may clarify the Hamiltonian in our model which may play an important role in the intriguing result in this section. From Eqs. (2)-(4), when we set $|\,h_A\,| = |\,h_B\,| = h$, the interacting Hamiltonian describing P-1 (AP-1) system may be given as

$$\hat{H}_{\text{int},NM} = -\eta\sigma\Delta_{SO}\tau^z, \ \hat{H}_{\text{int},FM1} = -\left[\eta\sigma\Delta_{SO} - \Delta_E + (+)\sigma h\right]\tau^z - \mu,$$

$$\hat{H}_{\text{int},FM2} = -\left[\eta\sigma\Delta_{SO} - \Delta_E + (-)\sigma h\right]\tau^z - \mu, \quad (14)$$

and that for P-2 (AP-2) system may be given as

$$\hat{H}_{\text{int},NM} = -\eta\sigma\Delta_{SO}\tau^z, \ \hat{H}_{\text{int},FM1} = -\left[\eta\sigma\Delta_{SO} - \Delta_E\right]\tau^z - \left[\mu + (+)\sigma h\right],$$

$$\hat{H}_{\text{int},FM2} = -\left[\eta\sigma\Delta_{SO} - \Delta_E\right]\tau^z - \left[\mu + (-)\sigma h\right], \quad (15)$$

where $\hat{H} = v_F\left(\hat{p}_x\tau^x - \eta\hat{p}_y\tau^y\right) + \hat{H}_{\text{int}}(x)$.

We first study the transmission probability ($T_{\eta\sigma}$) as a function of angle of incidence ($\theta$). Fig. 3 shows the transmission rates when $\Delta_E$ =$\mu$=U=0. The effect of exchange field on transmission is solely investigated. The result shows that the transmission rates at P-junctions seen in Figs. 3(a) and (c) are formed into two groups. In P-1 junction, we get $T_{P-1,k\uparrow} = T_{P-1,k\downarrow}$ and $T_{P-1,k'\uparrow} = T_{P-1,k'\downarrow}$, yielding "valley-polarization (VP $\neq$ 0)" with neither spin-polarization (SP=0) nor lattice-pseydospin polarization (LSP=0). This result may be described using Eq. (14). At ferromagnetic barriers, the energy gap of electrons in $k$ -valley with spin $\sigma$ is $E_{G,k\sigma} = 2(\Delta_{so} + h) \cong 18$ meV, while that of electrons in $k'$ -valley with spin $\sigma$ is $E_{G,k'\sigma} = 2(\Delta_{so} - h) \cong 2$ meV. At the same Fermi level $\mu = 0$, small energy gap may be able to conduct electric current better than high energy gap, giving rise to $T_{P-1,k'\sigma} > T_{P-1,k\sigma}$. In contrast, the transmission rates at P-2 junction shown in Fig. 3(c) yields $T_{P-2,k\uparrow} = T_{P-2,k'\uparrow}$ and $T_{P-2,k\downarrow} = T_{P-2,k'\downarrow}$, leading to SP $\neq$ 0 while VP=0 and LSP=0. This result may be described differently from that in P-1 junction, because in this case $E_{G,k'\sigma} = E_{G,k\sigma} = \Delta_{so}$. Using Eq. (15), the exchange field in this configuration may split the energy level $E + (-)h$ of electrons with spin $\uparrow (\downarrow)$ around the Dirac point ( E = 0 ). Therefore, Electrons with spin $\uparrow$ occupy energy level at $E + h \cong 9$ meV, which is greater than that of electrons with spin $\downarrow$ occupying energy level at $E - h \cong -1$ meV. For the same energy gap, higher energy level may give rise to higher transmission $T_{P-2,k'(k)\uparrow} > T_{P-2,k'(k)\downarrow}$, as seen in Fig. 3(c). This result may point out that the parallel junction may be either a pure valley filter device or a pure spin filter device, which is selectable by changing the direction of exchange fields. In AP-junctions shown in Figs. 3(b) and (d), we get VP=SP=LSP=0, where $T_{AP,k\uparrow} = T_{AP,k\downarrow} = T_{AP,k\uparrow} = T_{AP,k\downarrow}$. This is due to asymmetric configuration of AP-1 and AP-2 to induce the same net transmission when electron moving through NM1, FM1, NM2, FM2 and NM3. This result implies that asymmetric dual-ferromagnetic-barrier may break spin-valley filtering, see Figs. 3(b) and (d). On the contrary, breaking of spin-valley filtering effect by changing magnetic directions may not be achievable in single-barrier ferromagnetic gated silicene junction [21, 24].

Fig. 4 shows the transmission probability when $\Delta_E$ =4 meV, $\mu$=2.5 meV and U=0. The spin-valley and lattice-pseudospin polarizations are predicted when $T_{AP,k'\uparrow} \neq T_{AP,k'\downarrow} \neq T_{AP,k\uparrow} \neq T_{AP,k\downarrow}$ in parallel cases seen in Fig. 4(a) and (c). In P-1 junction, using Eq. (14), we get $E_{G,\eta\sigma} = 2|\eta\sigma\Delta_{SO} - \Delta_E + \sigma h|$, giving rise to $E_{G,k\uparrow} \neq E_{G,k\downarrow} \neq E_{G,k'\uparrow} \neq E_{G,k'\downarrow}$. The transmission probability in AP-junctions seen in Figs. 4(b) and (d) exhibits neither spin nor valley polarizations (SP=VP=0). Interestingly, the currents are split into two groups of lattice-pseudospin states $(T_{AP,k\downarrow} = T_{AP,k'\uparrow}) \neq (T_{AP,k\uparrow} = T_{AP,k'\downarrow})$ giving rise to pure lattice-pseudospin polarization $LSP \neq 0$, which is induced by interplay of $\mu$ and $E_Z$. In AP-junctions, asymmetric barriers may lead to the transmission that exhibits anisotropic tunneling like being influenced by a magnetic vector potential as seen in Figs. 4(b) and (d). Such anisotropic tunneling in angular space has been reported in gapped graphene based pseudospin valve [35] when the junction is anti-parallel.

Fig. 5 shows the spin-valley conductance as a function of applied electric field $E_Z$ when $\mu = U = 0$. In Fig. 5(a), perfect spin filter $SP \cong \pm 100\%$ occurs when $eDE_Z = \pm h$. Pure spin current can be controlled by reversing direction of electric field in P-1 junction. Interestingly, it is consistent with the result in Fig. 3(a) that pure valley polarization occurs $(G_{k'\uparrow} = G_{k'\downarrow}) \neq (G_{k\uparrow} = G_{k\downarrow})$ at $eDE_Z = 0$. It may be said that P-1 junction is a perfect spin filter with $SP \cong \pm 100\%$ at $eDE_Z = \pm h$, while it may turn into a valley polarizer at $eDE_Z = 0$. In contrast to the behavior at P-1 junction, perfect valley filter $VP \cong \pm 100\%$ occurs when $eDE_Z = \pm h$ at P-2 junction as seen in Fig. 5(c). It is also interesting to see consistency with the result in Fig. 3(c) that pure spin polarization occurs when $eDE_Z = 0$. In summary, P-junction is a perfect valley filter with $VP \cong \pm 100\%$ at $eDE_Z = \pm h$, while it may turn into a spin polarizer at $eDE_Z = 0$. In Figs. 5(b) and (d), both spin and valley polarization disappear (SP=VP=0), while a weak lattice-pseudospin filter occurs when varying $E_z$ around zero.

Figs. 6 (a) and (c) show perfect spin-valley filtering in P junctions, when gate potential $\mu = 2$ meV is applied into the ferromagnetic barriers. This result is similar to what was reported previously in a single ferromagnetic gated junction [24]. The distance between peaks may be adjusted by varying $\mu$. Stronger lattice-pseudospin filtering can be achieved by applying $\mu$ as seen in Figs. 6(b) and (d). Furthermore, the resolution of peaks splitting in Fig. 6 is enhanced by applying $U = 100$ meV as illustrated in Fig. 7. This result is significant for perfect spin-valley filtering and lattice-pseudospin filtering devices based on silicene junctions.

As we have discussed above, the P-1 and P-2 junctions can be considered as a pure valley polarizer and pure spin polarizer. In Fig. 8, we investigate polarizations controllable by gate potentials $\mu$ and U in P-1 and P-2 junctions. It is found that pure valley polarization ($SP = LSP = 0$ but $VP \neq 0$) in P-1 junction can be switched from almost $-100\%$ to $+100\%$ controlled by increasingly varying $\mu$ or U (see Fig. 8(a) and (c)). Also, pure spin polarization ( $SP \neq 0$ but $LSP = VP = 0$ ) in P-2 junction can be switched from almost +100% to −100% controlled by increasingly varying μ or U (see Fig. 8(b) and (d)). Finally, a pure lattice-pseudospin polarizer using AP-junction is investigated as a function of electric field $eDE_z$ (see Fig. 9 ) for $\mu = 2.5\,\mathrm{meV}$, where we focus on the effect of U. We find that at U=100, lattice-pseudospin polarization in both AP-1 and AP-2 junctions may be perfectly and linearly controlled from -100% to +100% by varying electric field. This is a good characteristic which may find applications in silicene-based lattice-pseudospintronics.

## 6. SUMMARY AND CONCLUSION

We have studied transport properties in silicene-based normal/ferromagnetic/ normal/ferromagnetic/normal junction, controlled by dual ferromagnetic-gated barriers. We focused on the effect of exchange fields, which are induced into two magnetic regions, on transport properties. As a result, the parallel junctions exhibit interesting behaviors that they can be either pure spin polarizer or pure valley polarizer depending on the direction of exchange fields. The perfect spin-valley filtering effect is also found in the parallel junctions similarly to that in single ferromagnetic-gated junctions [24]. As an interesting behavior, the results showed that, the asymmetric barriers of anti-parallel junctions may break spin-valley filtering effect to yield a pure lattice-pseudospin polarization, which can be controlled linearly from -100% to +100% by varying electric field. This is to say that the current in anti-parallel junctions can be controlled to flow only in A-or B-sublattice when LSP= +100% or -100%, respectively. The perfectly controllable lattice-pseudospin current in silicene was found only in double ferromagnetic-gated anti-parallel junctions. We expect our work to be verified and make contribution to the field of not only silicene-based spin-valleytronics, but also lattice-pseudospintronics.

## ACKNOWLEDGMENTS

This work was supported by Kasetsart University Research and Development Institute (KURDI) and Thailand Research Fund (TRF) under Grant. No.TRG5780274.

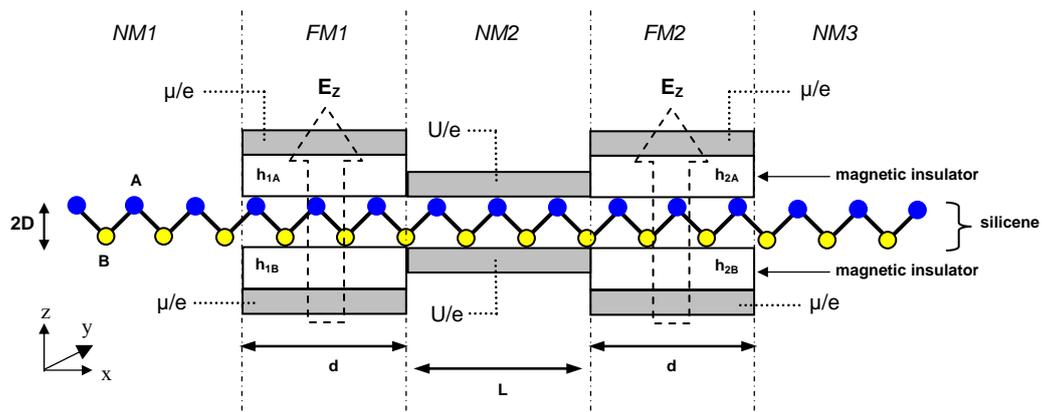

Fig. 1. Cross-sectional schematic model of double-barrier silicene-based NM1 / FM1 / NM2 / FM2 / NM3 structure. Electric field $E_z$ and gate potential $\mu / e$ are applied into the magnetic barriers, while gate potential $U / e$ is applied at the silicene NM2 layer.

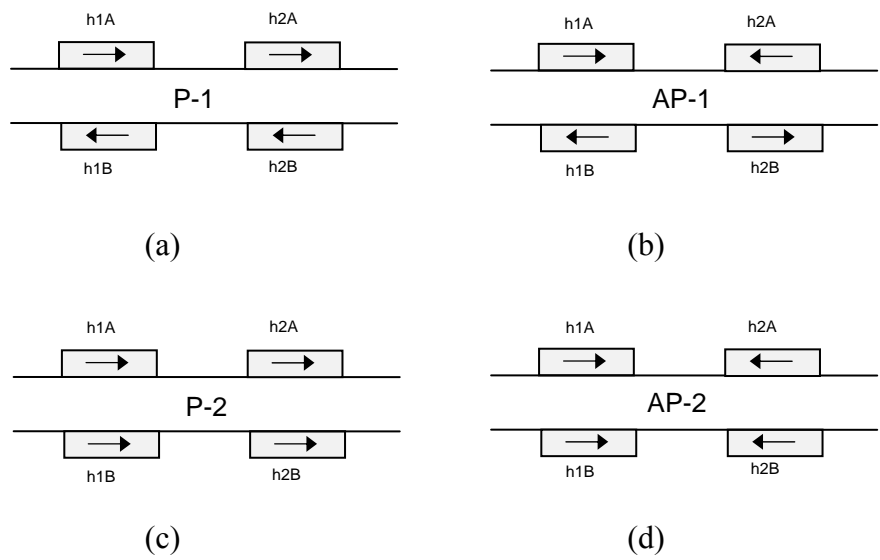

Fig. 2. Junction types used in this investigation, where → represents h, and ← represents –h. (a) Parallel junction type 1 (P-1), (b) anti-parallel junction type 1 (AP-1), (c) parallel junction type 2 (P-2), (d) anti-parallel junction type 2 (AP-2).

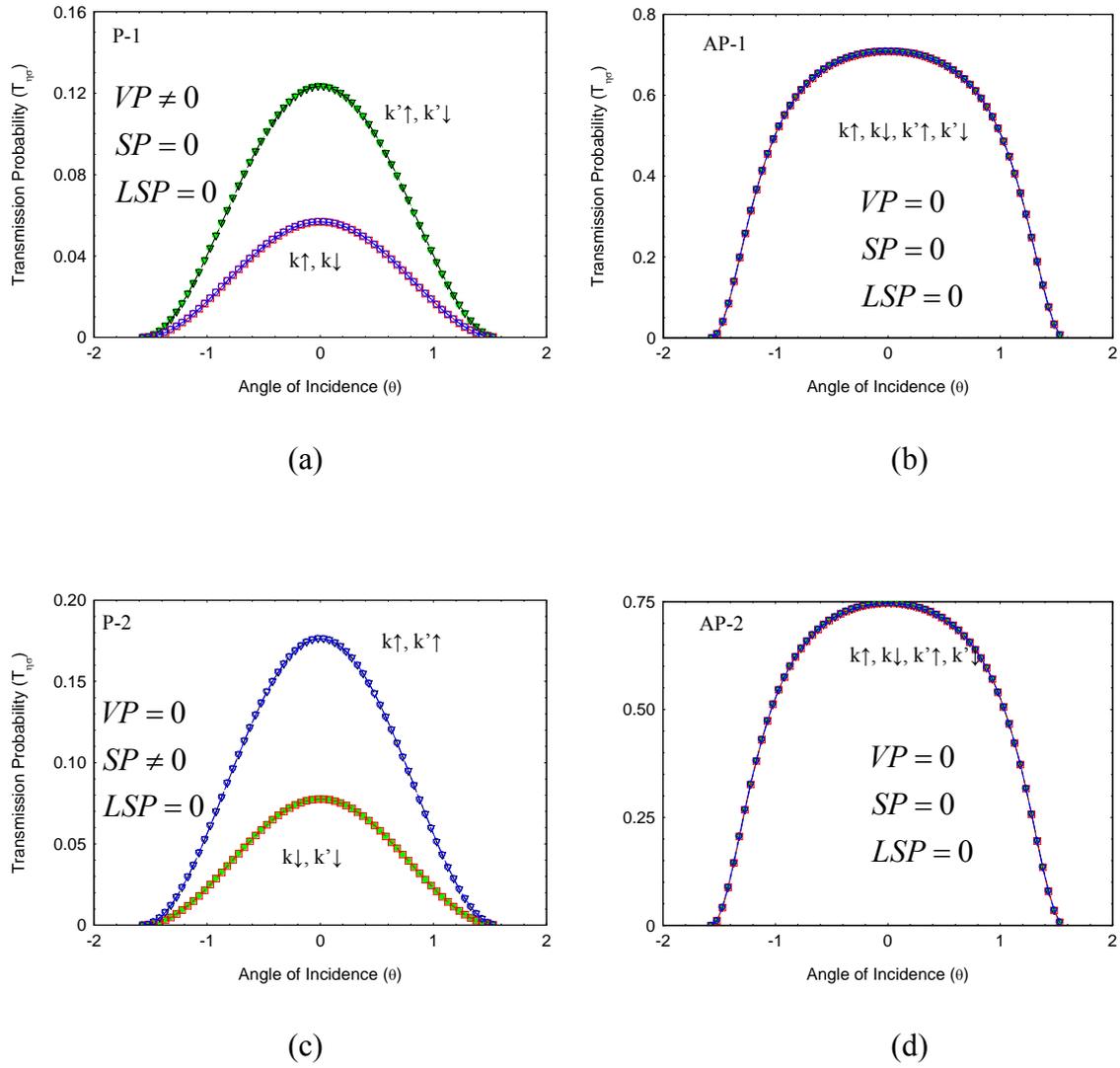

(a)

(b)

(c)

(d)

Fig. 3. Transmission rates as the function of θ in P and AP junctions when L=25 nm, d=25 nm, E=4 meV, h=5 meV, $eDE_Z$=0 meV, μ=0 meV, and U=0 meV. (a) In P-1 junction, (b) in AP-1 junction, (c) in P-2 junction, and (d) in AP-2 junction.

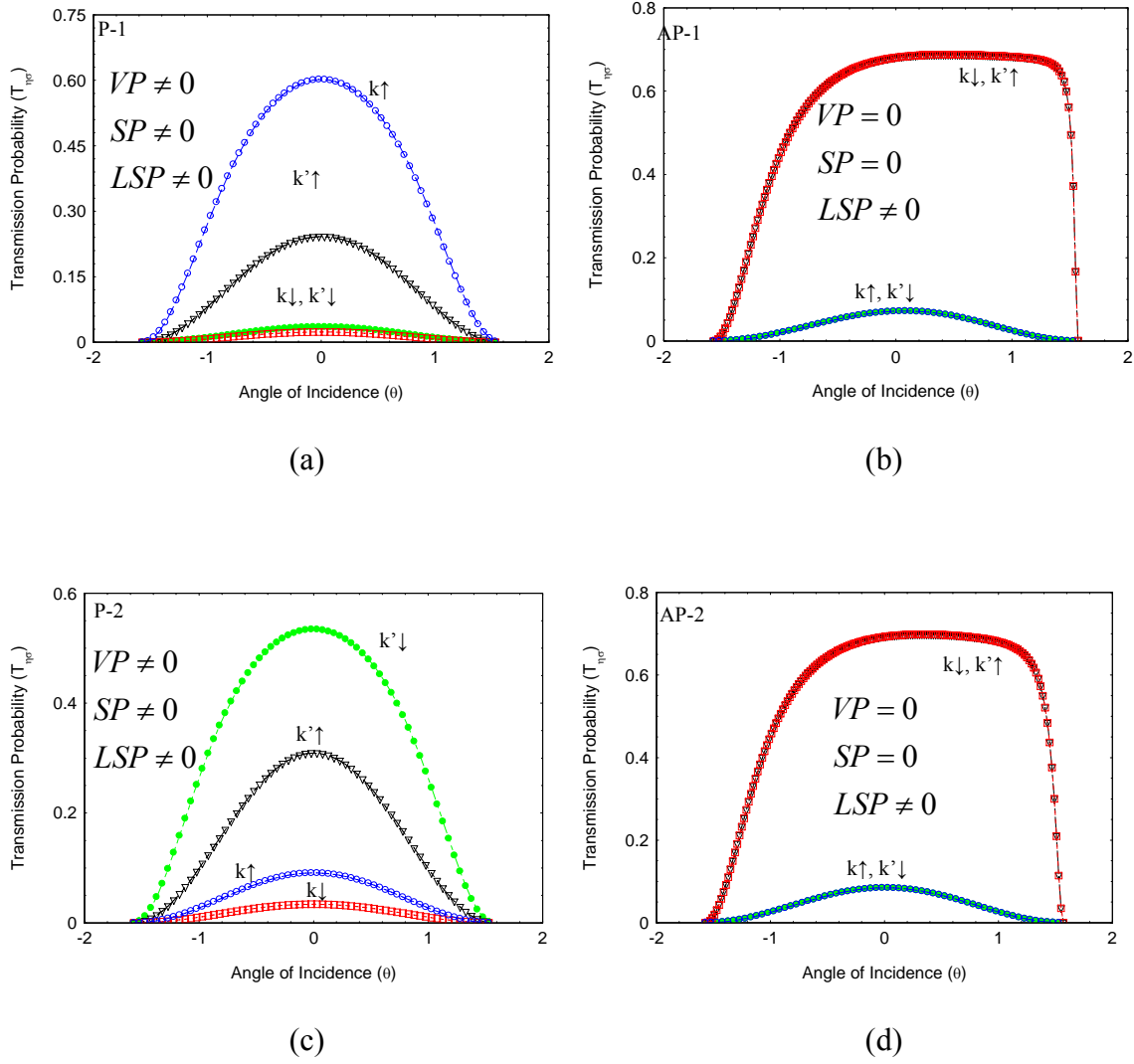

Fig. 4. Transmission rates as the function of θ in P and AP junctions when L=25 nm, d=25 nm, E=4 meV, h=5 meV, eDE$_Z$ =4 meV, μ =2.5 meV, and U=0 meV. (a) In P-1 junction, (b) in AP-1 junction, (c) in P-2 junction, and (d) in AP-2 junction.

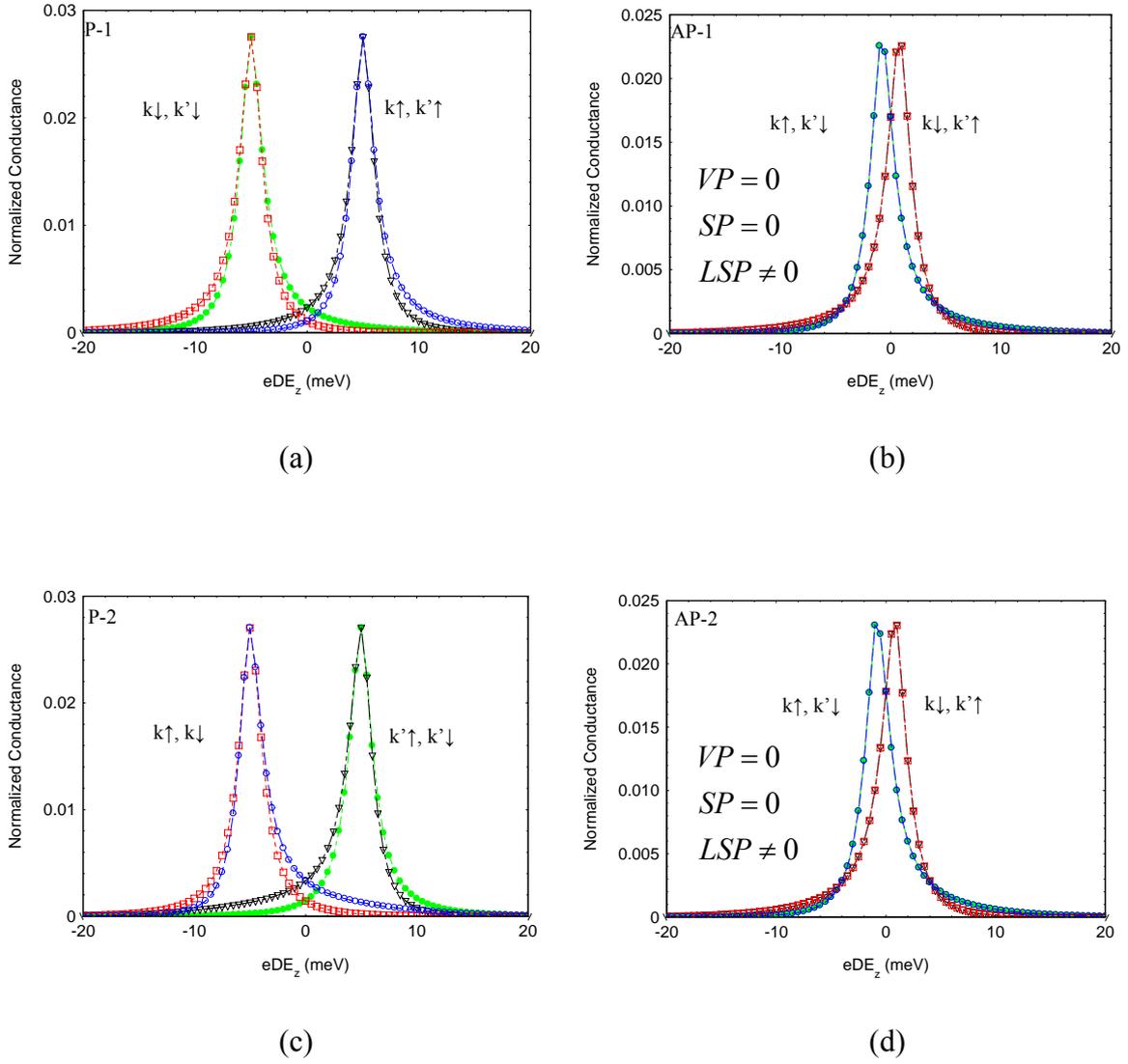

(a)

(b)

(c)

(d)

Fig. 5. Conductance as the function of eDE$_z$ in P and AP junctions when L=25 nm, d=25 nm, E=4 meV, h=5 meV, μ=0 meV, and U=0 meV. (a) In P-1 junction, (b) in AP-1 junction, (c) in P-2 junction, and (d) in AP-2 junction.

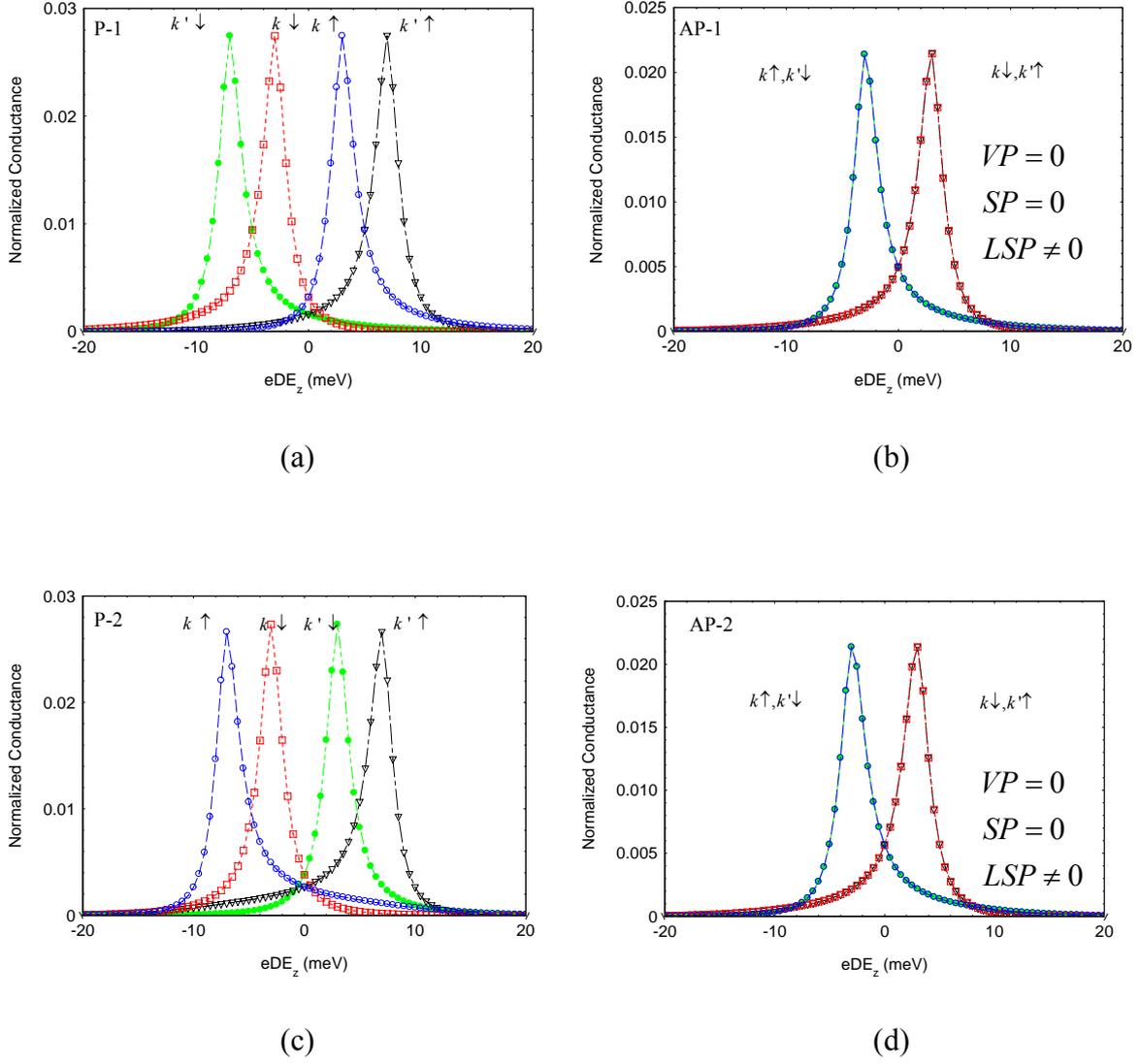

(a)

(b)

(c)

(d)

Fig. 6. Conductance as the function of $eDE_z$ in P and AP junctions when L=25 nm, d=25 nm, E=4 meV, h=5 meV, μ=2.0meV, and U=0meV. (a) In P-1 junction, (b) in AP-1 junction, (c) in P-2 junction, and (d) in AP-2 junction.

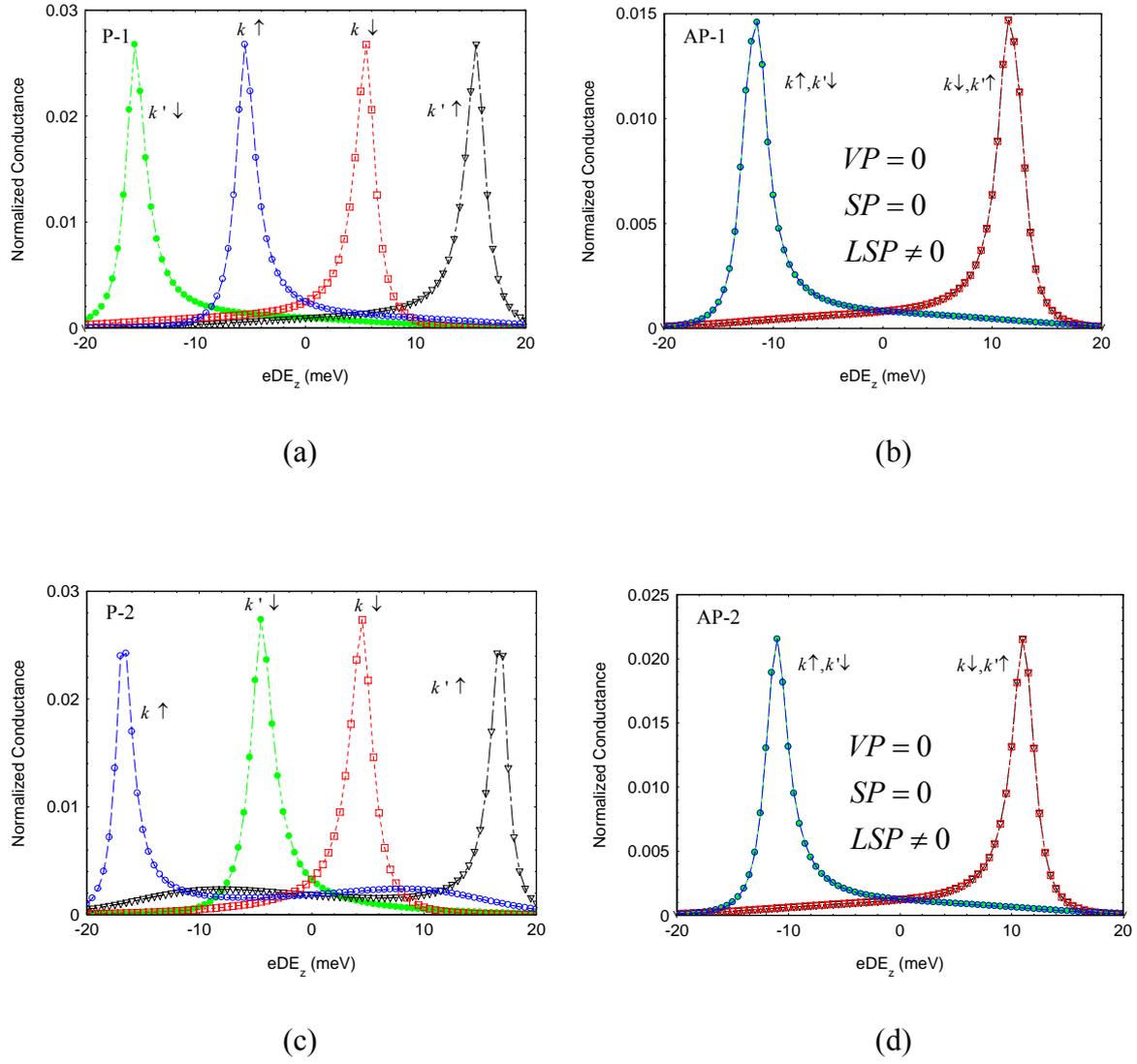

(a)

(b)

(c)

(d)

Fig. 7.  Conductance as the function of $eDE_z$ in P and AP junctions when L=25 nm, d=25 nm, E=4 meV, h=5 meV, μ=2.0 meV, and U=100 meV.  (a) In P-1 junction, (b) in AP-1 junction, (c) in P-2 junction, and (d) in AP-2 junction.

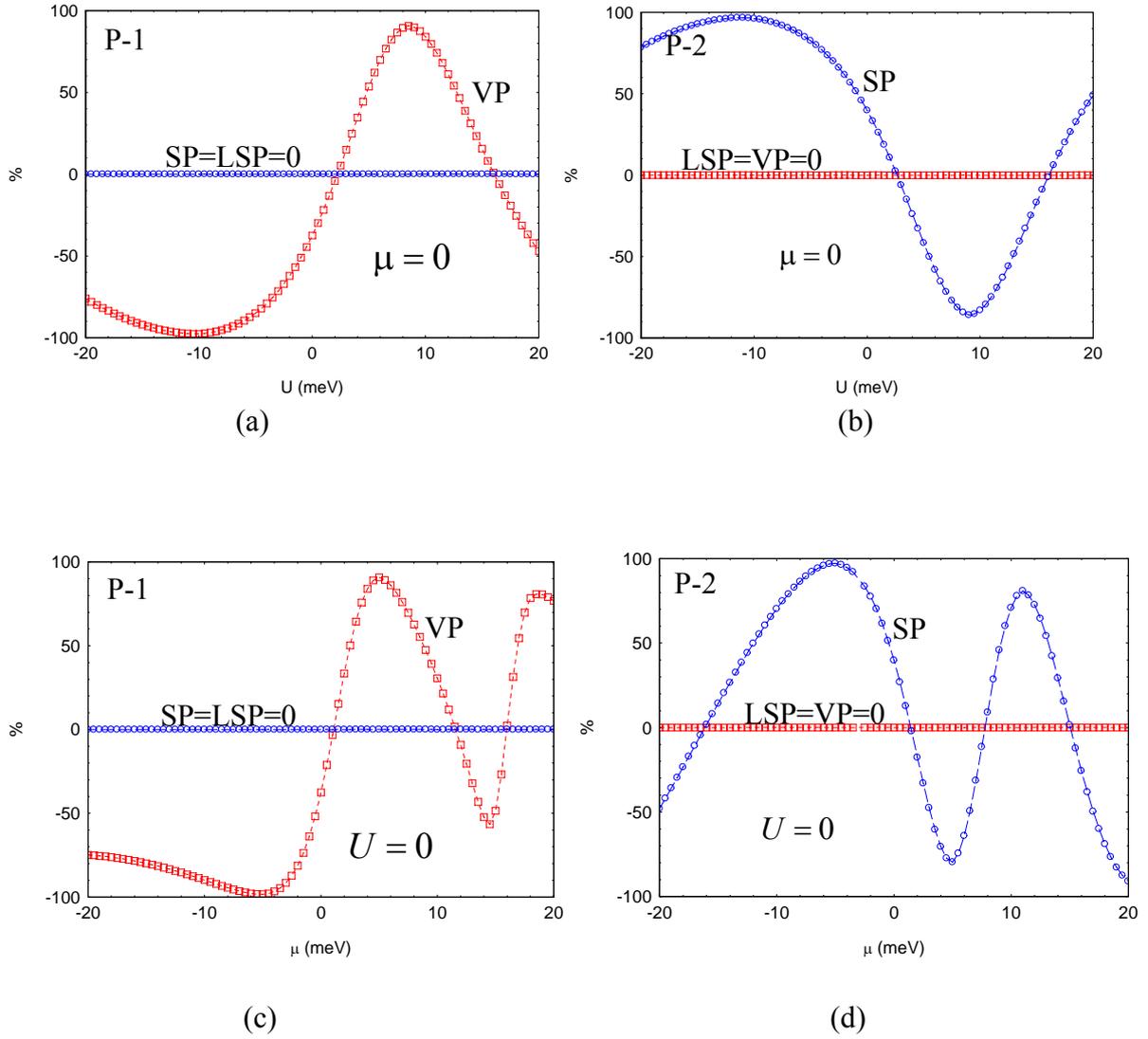

Fig. 8. Spin polarization (SP) and valley polarization (VP) as the function of U and μ in P junction when L=25 nm, d=25 nm, E=4 meV, h=5 meV, and E$_z$=0 meV. (a) In P-1 junction and (b) in P-2 junction as a function of U with μ=0meV. (c) In P-1 junction and (d) in P-2 junction as the function of μ with U=0 meV. Our junction can be either a pure spin polarizer or pure valley polarizer by selecting suitable magnetic direction.

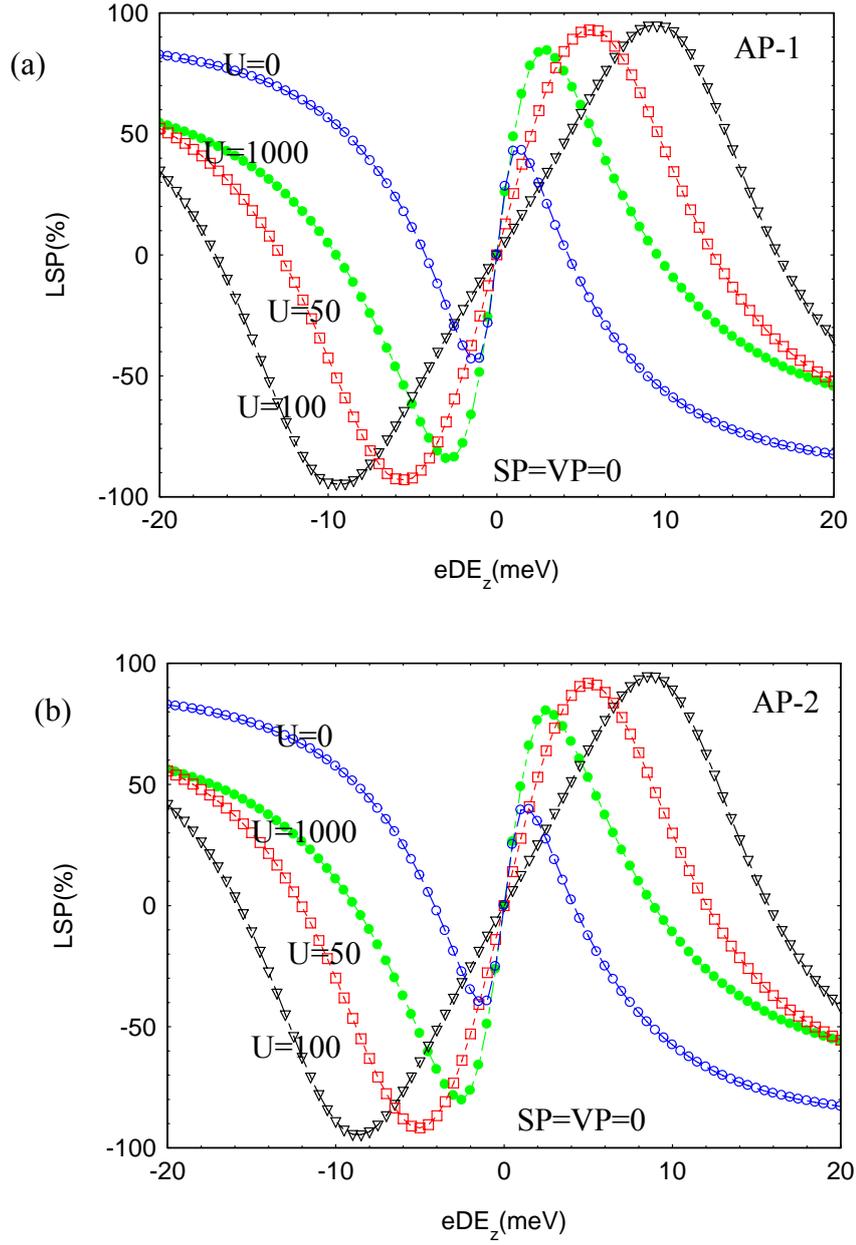

Fig. 9. Lattice-pseudospin polarization (LSP%) L=25 nm, d=25 nm, E=4 meV, h=5 meV, μ=2.5 meV. (a) In AP-1 junction and (b) in AP-2 junction when U=0, 50, 100, 1000 meV.